\documentclass[12pt]{article}

\usepackage{graphicx}
\usepackage{dcolumn}
\usepackage{bm}

\begin{document}

\title{Theory of the Observed Ultra-Low Friction between Sliding Polyelectrolyte Brushes}

\vskip 1cm
\author{ J. B. Sokoloff\\ Department of Physics and\\Center for Interdisciplinary Research on Complex Systems\\
 Northeastern University, Boston, Massachusetts 02115, U.S.A., \\E-mail address: j.sokoloff@neu.edu}

\maketitle

\begin{abstract}
It is shown using a method based on a modified version of the mean field theory of Miklavic Marcelja 
that it should be possible for osmotic pressure due to the counterions associated with the two polyelectrolyte polymer brush coated surfaces to support a reasonable load (i.e., about $10^6 Pa$)
with the brushes held sufficiently far apart to prevent entanglement of polymers belonging to the two brushes, thus avoiding what is believed to be the dominant mechanisms for static and dry friction. 
\end{abstract}


\section{Introduction}

Polymer brush coatings on solid surfaces provide very effective lubrication, 
in the sense that they are able to support significant load (pushing the surfaces 
together), but have exceedingly low friction coefficients\cite{klein}. Human and animal joints 
are known to exhibit very low friction and wear. The outer surface of the cartilage coating 
these joints have polymeric molecules protruding from them\cite{McCrutchen,schartz,dowson,McCrutchen1,maroudas,mow,murukami,ateshian,
kirk,McCrutchen2,swann,jay,schumacher}. This suggests 
the strong possibility that their very effective lubrication is a result of 
polymer brush lubrication.  
It has been shown that at small loads, polymer brush coated 
surfaces can slide relative to each other with the bulk of the brushes not 
in contact\cite{sokoloff}. If the surfaces are far enough apart under such loads, the load will be supported almost entirely by osmotic pressure due to a dilute concentration of polymers that protrude into the thin interface 
region separating the brushes. As a consequence, there will be little entanglement of the polymers 
belonging to the two brushes (i.e., penetration of polymers belonging to one brush into the second brush). It was argued in Ref. \cite{sokoloff} that such entanglement of polymer brushes leads to static and kinetic friction that saturates at a nonzero value in the limit of zero sliding velocity (i.e., dry friction) of polymer brush coated surfaces. It is reasonable to expect that there should 
be little wear as well when there is little entanglement.

The discussion in Ref. \cite {sokoloff} deals with static and slow speed kinetic friction, and not the purely viscous kinetic friction that occurs at high sliding speeds. Since the polymers hyaloronan or lubricin, which coat the cartilage 
in human and animal joints\cite{ruths}, are charged, it is necessary 
to consider polyelectrolyte brushes, whose equilibrium properties have 
been studied using mean field theory by Zhulina, et. al.\cite{zulina}, Misra\cite{misra} 
and Miklavic\cite{miklavic}, and Pincus and Tamashiro, et. al.\cite{pincus}. Raviv, et. al.\cite{raviv}, have found that polyelectrolyte brushes exhibit remarkably low friction coefficients ($10^{-3}$ or less) compared to the friction coefficients typically found for neutral polymer brushes\cite{klein,kumacheva,fetters}. It is proposed in the present article that 
polyelectrolyte polymer brushes are more effective lubricants than 
neutral polymer brushes discussed above because, as will be shown, counterion osmotic pressure present in a relatively thin interface region separating two polymer brushes is able to support a load of the order of $10^6 Pa$ (which is comparable to the loads supported by the polyelectrolyte brushes studied in Ref. \cite{raviv}), without the tops of the bulk part of the density profiles of the brushes being in contact, as illustrated in Fig. 1. 
\begin{figure}
\center{\includegraphics [angle=0,width=10cm]{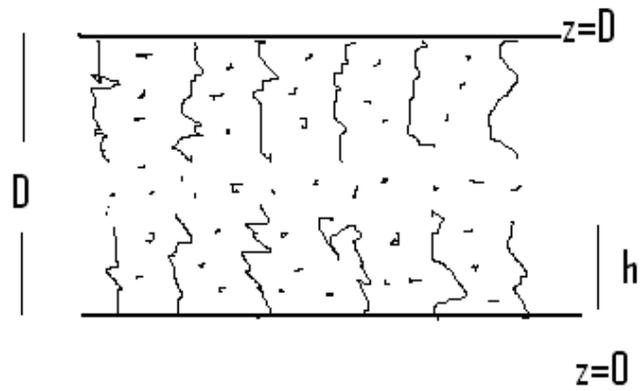}}
\caption{The geometry of two polyelectrolyte  polymer brush coated surfaces with the load pushing the surfaces together supported by osmotic pressure due to counterions in the interface regions separating the tops of the brushes is illustrated schematically. The dots located among the polymer chains and in the interface region between the two brushes represent the counterions. As illustrated here, D denotes the spacing of the surfaces and h denotes the polymer brush height.}
\label{fig4}
\end{figure}
The height of a neutral polymer brush is determined by a competition between the 
polymer's elasticity (of entropic origin) and the mutual repulsion that occurs in 
a good solvent between the monomers making up the polymers of the brush.  
For polyelectrolyte brushes, the counterion entropy\cite{pincus} also plays a significant role in stretching the polymers in the brush. 

The present treatment of lubrication due to polyelectrolyte brushes is based on the theory of Miklavic and Marcelja\cite{miklavic}, which is an extension of the analytic mean field theory due to Milner, et. al.\cite{milner}. As has been shown in several treatments of the subject\cite{kim}, using numerical treatments of mean field theory and molecular dynamics\cite{grest}, the analytic treatment of mean field theory\cite{milner} is only accurate for relatively stiff polymer brushes, for which the brush height is large compared to the unswollen single polymer radius of gyration $N^{1/2}a$, where a is the monomer length. Since for good solvents this is a useful limit to consider, a treatment based on analytic mean field theory is expected to give accurate results in the important case of relatively dense polymer brushes with relatively strong repulsions between their monomers, as well as for polyelectrolytes for which counterion osmotic pressure within the brushes provides strong forces to keep the polymers in the brush well stretched\cite{pincus}.  Furthermore, analytic mean field theory shows important trends as the brushes are compressed\cite{sokoloff}, which are consistent with experiment. Therefore, a theoretical study of lubrication by polymer brushes based on mean field theory is expected to be a good starting point in an effort to understand lubricating properties of polymer brushes. In section 2, it is argued that the method of estimating the mean distance that a polymer belonging to one of the brushes is able to extend into the interface region separating the two brushes, used for neutral polymer brushes, applies equally well to polyelectrolyte brushes, when they are sufficiently compressed, and this distance is estimated.  If the interface region separating the tops of the brushes is wider than this distance, polymers from one brush will be unlikely to penetrate into the second. In section 3, the question of how much load can be supported by counterion osmotic pressure (with the brushes held sufficiently far apart to prevent the occurrence of static friction) is studied. 

In an effort to determine a possible mechanism for the ultra-low friction found in Ref. \cite{raviv}, an approximate solution for the Poisson-Boltzmann equation beyond the Debye-Huckel approximation will be used to determine the concentration of counterions in a region located midway between the two polyelectrolyte brushes. This result is used to show that for polyelectrolyte brushes, osmotic pressure due to the counterions is capable of supporting about $10^6 Pa$ of load, even with the tops of the mean field theoretic monomer profiles of the two brushes about $14A^o$ apart. In fact, it will be shown that this result is valid even for salt concentrations smaller than but still comparable to 0.1 M (about $10^{26} m^{-3}$), the salt concentration in living matter. Increasing the salt concentration further, however, will put the system in a regime in which Debye-Huckel (D-H) approximation is accurate and predicts that the ion concentration mid-way between the brushes is negligibly small once the D-H screening length is much smaller than the width of the interface region.

\section{Estimate of the Brush spacing for which Static Friction due to Brush Entanglement will Not Occur}

Part of Ref. \cite{sokoloff} relevant to the present discussion is briefly summarized in this paragraph. 
Milner, et. al.\cite{milner}, proposed a simple way to solve the mean 
field theory analytically for neutral polymers attached at one end to a surface. In this treatment, 
the location of the $n^{th}$ monomer of the $i^{th}$ polymer belonging to a polymer brush of minimum free energy, 
${\bf r_i (n)}$ satisfies the differential equation
$${d^2 {\bf r_i (n)}\over dn^2}=\nabla V({\bf r_i (n)}), \eqno (1)$$
where $V({\bf r_i (n)})=w\phi ({\bf r_i (n)})$, and 
where for neutral brushes, $\phi ({\bf r})$ is the monomer number density and w is the strength of 
the monomer-monomer repulsion parameter. This can be thought of as an "equation 
of motion" for the monomers, in which the index n labeling the monomers formally 
plays the role of time. It is formally analogous to Newton's second law for motion 
of a particle in a potential equal to $V({\bf r (n)})$. Existence of a solution 
of the mean field equations of motion requires that $V({\bf r (n)})$ be a parabolic 
function of z \cite{milner}. In Ref. \cite{sokoloff} it was shown that when two polymer brushes are in contact with each other or nearly in contact, polymers belonging to one brush can penetrate into the second brush. As a consequence, there is a force of static friction equal to the force needed to pull these intertangled polymers out. The force of static friction per intertangled polymer was argued to be equal to $k_B T/\xi$, where $\xi$ is the mesh size of one of the two polymer brushes. The magnitude of the friction was estimated in Ref. \cite{sokoloff} to be of the order of $10^3 Pa$ or more. Even when the applied force is below this value, the surfaces will not remain truly stationary, but rather, will creep relative to each other. The reason for this behavior is that the intertangled polymers will diffuse out in a diffusion or reptation time $\tau$. It was shown that at high compression $\tau$ can be sufficiently long to consider this to be a true force of static friction. If a force greater than this static friction force is applied, and hence, the surfaces slide with a speed much greater than this creep speed, since the polymers will no longer have enough time to re-entangle in the second brush, the friction will no longer be determined by this mechanism. Instead, there will be viscous friction resulting from the fact that the solvent gets sheared as the surfaces slide. This viscous friction force per unit surface area in the slow sliding speed limit is given by $\eta (v/\ell_p)$, where $\eta$ is the viscosity of the solvent, v is the sliding velocity\cite{deGennes1} and $\ell_p$ is the hydrodynamic penetration length into the brushes.\cite{milner1}. Let us consider sliding speeds that are sufficiently slow so that the zero sliding speed configuration of the polymers is not significantly disturbed. For uncompressed polymer brushes, which have a parabolic density profile\cite{milner}, $\ell_p$ is comparable to the equilibrium brush height\cite{milner1}, but for brushes which are compressed because they are supporting a load, the density profile will get flattened out, and as a consequence, $\ell_p$ can be comparable to the polymer spacing at the surface to which they are attached, s. Then, for example, assuming the solvent to be water, for v=1 mm/s, the largest speed used in the experiment, the force of friction per unit area for the uncompressed brush case will be approximately equal to about a $Pa$, for brushes of equilibrium height of the order of $500 A^o$. For compressed brushes with an anchor spacing s of $84 A^o$, it will be about 5 times this value.  
This estimate is an upper bound on the shear stress, since it assumes no-slip boundary conditions for the fluid at the polymers and the surface. For plate spacing just above the maximum separation for which the friction observed is above the experimental accuracy (reported in Ref. \cite{fetters}), the load is reported to be F/R=0.01N/m (where R is the radius of the cylinders in the surface force apparatus, which is about 0.01 m), which, using the standard Hertz formula\cite{johnson}, gives a contact area for the surface force apparatus of about $10^{-10}m^{-2}$. The maximum observable shear force in this experiment is $0.25\mu N$, which when divided by the above contact area gives a shear stress from the experimental data of about $10^4 Pa$, clearly well above the shear stress of  about a $Pa$ found above for viscous friction. Hence the viscous friction discussed above is clearly well below the experimentally observed kinetic friction. In contrast, Eq. (16) in Ref. \cite{sokoloff} shows that it is easy to get a value for the static friction or slow speed kinetic friction, due to the polymer blob entanglement mechanism discussed earlier, comparable or greater than the experimental value if the brushes are sufficiently compressed.

An important advantage of 
polyelectrolyte over neutral polymer brushes as lubricants is that the osmotic pressure due to counterions might, under 
the right conditions be able to support the load, allowing the brush coated surfaces to slide 
without the bulk of the brushes being in actual contact, and hence, with negligible friction due to entanglement of polymers from one brush in the second.
In order to calculate the separation of the two brushes above which friction due to this entanglement no longer occurs and to calculate the contribution to the repulsion of two polyelectrolyte brushes due to osmotic pressure due to the counterions at this separation, we will use an analytic mean field theory treatment  
of polyelectrolyte brushes based on the treatment due to Miklavic and Marcelja\cite{miklavic}. In their 
treatment they use the mean field treatment of Ref. \cite{milner} with $V({\bf r})=w\phi ({\bf r})+e\psi ({\bf r}),$ where $\psi({\bf r})$ is the electrostatic potential, due to the 
charged polymers making up the brush, screened by the counterions divided by the electronic charge e, which satisfies Poisson's equation 
$${d^2\psi\over dz^2}=4\pi\rho (z)/e-(4\pi f/\epsilon)\phi (z), \eqno (2)$$
where $\rho (z)$ is the ionic charge density, z is the distance from the lower surface, $\phi (z)$ is the monomer density profile 
of the brush (or brushes, if there are two of them in contact or nearly in contact) and f is the charge in units of the electron charge e per monomer. In contrast to Ref. \cite{miklavic}, in the present treatment of this problem, the Debye-Huckel approximation will not be used when estimates of the osmotic pressure due to counterions are made. The charge per monomer f has a maximum value because of Manning condensation\cite{manning}.  If the charge spacing on the polymer 
is less than the Bjerrum length, there will be Manning condensation\cite{manning}, which means that some of the counterions 
will condense onto the polymer, until its charge density is reduced to the point that the charge spacing becomes equal to 
a Bjerrum length for monovalent counterions. Then, since the Bjerrum length, $\ell_B=e^2/(\epsilon k_B T)$ is 7.1 Angstroms for a solvent with a dielectric constant comparable to that of water, the largest possible value of f is the ratio of a monomer spacing to a Bjerrum length, or 0.35, if we assume the value for the monomer spacing from Ref. \cite{raviv} of $2.5A^o$. Since we want the polymer brushes to behave as polymer brushes in a good solvent, we want the charge per monomer to be sufficiently small so that there are not too many counterions condensed on the polymers, in order to prevent possible collapse of the brushes due to interaction of dipole moments resulting from counterions condensed on the charged monomers\cite{schiessel}. 
Milner, et. al. \cite{milner} and Miklavic and Marcelja\cite{miklavic} show that in mean field theory the height of the $n^{th}$ monomer on a polymer in a brush, $z_n$ satisfies the differential equation
$$d^2z_n/dn^2=(a^2/k_B T)(\partial V(z_n)/\partial z_n), \eqno (3a)$$
where for polyelectrolyte brushes $V(z_n)$ is the potential acting on the $n^{th}$ monomer, due in the present case to both hard core intermonomer forces and electrostatic forces. In order to have a self-consistent solution to Eq. (3a), $V(z)$ must have the form $A-Bz^2$, where A and B are constants\cite{milner}. 

Let us now estimate effects of fluctuations from mean field theory by a method similar to the methods of Ref. \cite{milner}. 
If $j_1$ monomers belonging to a polymer are either pulled out or thermally fluctuate out 
of a brush, the solution for $z_n$ as a function of n becomes for $z_n\leq h$
$$z_n=(h/cos(\omega j_1))cos(\omega n),\eqno (3b)$$
using Eq. (5b) of Ref. \cite{sokoloff} and requiring that $z_{j_1}=h$, 
where $j_1$ is the value of n such that for $n<j_1$, $z_n>h$\cite{sokoloff}. For $z_n>h$ the "force" in the "equation of motion" is 
no longer determined by $-(A-Bz^2)$, but by the electrostatic potential $\psi (z)$ outside of the brush, which is a smooth function of z. Then for $n<j_1$, $z_n$ is given by 
$z_n=f(n)$, where $f(n)$ represents the solution to the "equation of motion" with the potential given by the value of $-\psi (z)$ outside the brush instead of $-(A-Bz^2)$. If $j_1<<N$ (a condition for the validity of mean field theory),  we may expand f(n) to lowest order in n for $n<j_1$. Hence, 
$$h=z_{j_1}=f(j_1)\approx f(0)+f'(0) j_1=z_0+f'(0) j_1,$$
or
$$z_0=h-f'(0) j_1. \eqno (4)$$
The assumption made here that $z_n$ is a slowly varying function of n in the region outside the brushes is valid since the distance over which the electrostatic potential varies is of the order of the thickness of the interface region separating the brushes and the distance that the polymers belonging to a brush stick out into this region is smaller than this distance, because we have chosen to consider the situation in which the spacing between the brushes is greater than the distance that the polymers belonging to a brush protrude into this region. We must require that $dz_n/dn$ be continuous at $n=j_1$. It follows from Eq. (3) that 
$${dz_n\over dn}|_{n=0}\approx f' (0)=-h\omega tan(\omega j_1)\approx -h\omega^2 j_1. \eqno (5)$$
This implies that $z_0\approx h+h\omega^2 j_1^2,$ and hence, $dz_0=2h\omega^2 j_1 dj_1$.
As in Ref. \cite{sokoloff}, we use the fact that $(k_B T/a^2)(dz_n/dn)|_{n=0}$ is the tension that must be 
applied to the free end of the polymer in order to pull it out of the brush to calculate the work 
needed to pull $j_1$ monomers of this polymer out of the brush,
$$\Delta F=((k_B T/a^2)\int_h^{z_0} (dz_n/dn)|_{n=0} dz_0=$$

$$(2/3)k_B T(h\omega^2/a)^2 j_1^3. \eqno (6)$$
From Eq. (5) we obtain the probability that $j_1$ monomers stick out of the polymer into the interface region, 
$$P\propto exp(-(2/3)(h\omega^2/a)^2 j_1^3) \eqno (7)$$ 
or using the fact that $h\omega^2j_1^2=z_0-h$ [which is given under Eq. (5)], we find that 
$$P\propto exp[-(\frac{z_0-h}{\xi})^{3/2}],  \eqno (8)$$
where $\xi/h=(3/\pi)^{2/3} (N^{1/2} a/h)^{4/3}.$ 
We expect that we will get extremely low friction, only if the polymers that fluctuate  out of one 
brush do not extend so far that they get entangled in the second polyelectrolyte brush, because that will result 
in static friction and relatively large nonviscous kinetic friction\cite{sokoloff}. For $D-2h=z_0-h>>\xi,$ there will be no static friction due to entanglement of polymers belonging to one brush in the second. Using the parameters N=115, $a=2.5A^o$ and $h=170A^o$ from Ref. \cite{raviv} in the formula for $\xi$ under Eq. (8), we find that we must have $z_0-h\geq \xi\approx 14.4 A^o$. In fact, numerical work on neutral polymer brushes\cite{kim,grest} shows that the monomer density is generally quite small in the tails on the monomer density distribution of length $\xi$, implied by Eq. (8), and there is every reason to assume that the same will be true for charged polymers as well, making it likely that it is not necessary for $z_0-h$ to be much larger than $\xi$ in order for the monomer density in the tails to be in the dilute regime\cite{sokoloff}, in which friction due to blob entanglement does not occur. 

The model discussed in this section assumes that the polyelectrolyte brushes have a height which is a relatively small fraction of the fully stretched length of the polymers, as is the case for neutral polymer brushes. This would be the case only for polyelectrolyte brushes for which the charge on the polymers is a very small fraction of the maximum charge allowed by Manning theory\cite{manning}, or for polymers with a higher charge but carrying a high load, so that the brush heights are compressed to a small fraction of the heights expected for uncompressed brushes on the basis of Pincus' arguments\cite{pincus}. For highly charged polyelectrolyte brushes under light loads, for which the polymers are extended to close to their fully extended length, the high degree of entanglement needed to give the static and relatively high kinetic friction discussed in Ref. \cite{sokoloff} for neutral brushes is unlikely. In fact, it was argued earlier that this is likely to be the reason that experimental studies of polymer brushes\cite{fetters} show that the friction is negligible until the brushes are under relatively high loads.     

\section{\bf Load Carrying Ability of Polyelectrolyte Brushes in the Low Salt Concentration Limit}

Since in the absence of excess salt, the conterion contribution to the osmotic pressure falls off quite slowly with plate separation\cite{pincus}, the possibility will be explored here that a reasonably large load can be supported by counterion osmotic pressure while the plates are sufficiently far apart to prevent entanglement of the polymers belonging to the two brushes.
Therefore, let us now consider the situation in which there is a sufficiently low concentration of excess salt. In this situation, we must use solutions of the Poisson-Boltzmann equation beyond the Debye-Huckel approximation, since that approximation does not accurately describe the problem\cite{pincus,safran}. Let us first consider the situation in which there is no excess salt, and only counterions are present. 
In the interface region (i.e., $h<z<D-h$), the electrostatic potential must be a solution of the Poisson-Boltzmann equation
$$d^2\psi/dz^2=-4\pi n_0\ell_B e^{\ell_B\psi},\eqno (9)$$
where $n_0$ is the counter-ion density midway between the plates, i.e., at z=D/2 \cite{safran}. The solution to Eq. (9) for this geometry in the interface region between the two polymer brushes may be used to estimate the counterion contribution to the osmotic pressure supporting the load. The two polymer brushes may be formally replaced by two equally charged flat plates a distance D-2h apart  if we impose the boundary condition that $-d\psi/dz|_{z=h}=\sigma_e$ as required by Gauss's law, where $\sigma_e$ is the total charge (polymer charge plus counterion charge) contained within the brush (i.e., in the region $0<z<h$). This solution is\cite{safran} 
$$\psi=\ell_B^{-1}log(cos^2 k_0 (z-D/2)),\eqno (10)$$
where $k_0^2=2\pi n_0\ell_B$, and the counterion density is given by
$$n=n_o e^{-\ell_B\psi}=\frac{n_0}{cos^2 k_0(z-D/2)}.\eqno (11)$$
Since
$$\int^{D-h}_h dz n(z)=n_0\int_h^{D-h}\frac{dz}{cos^2 k_0(z-D/2)}=$$

$$(2n_0/k_0)tan k_0(D/2-h)=2\sigma_e, \eqno (12)$$ 
$$tan k_0(D/2-h)=k_0\sigma_e/n_0, \eqno (13)$$
for $k_0\sigma_e/n_0>>1$, we get the maximum possible value of $k_0$, which gives the maximum value of $n_0$, namely, $k_0=[\pi/(D-2h)]$. From the definition of $k_0$ below Eq. (10), it follows that the largest possible value of 
$$n_0=(\pi/2)\frac{1}{\ell_B (D-2h)^2}, \eqno (14)$$
which gives for the counterion contribution to the osmotic pressure
$$P_{osm}=k_B T n_i =(\pi/2)\frac{k_B T}{\ell_B (D-2h)^2}. \eqno (15)$$
For a value of the parameter D-2h, comparable to $\xi$, estimated in section 2, or about 14.4$A^0$, we find that $P_{osm}=2.7\times 10^6 N/m^2$, and it is inversely proportional to the square of the spacing between the tops of the brushes, i.e., the width of the interface region.

In order to determine if the right hand side of Eq. (13) is much greater than one [which is the condition for the validity of Eqs. (14) and (15)], Eq. (2) will now be integrated, in order to determine whether a value of $n_0$ (and corresponding value for the osmotic pressure given above for the region separating the tops of the brushes) comparable to the value given above will occur. 
The Poisson-Boltzmann equation [Eq. (2)] for the case of no excess salt can be written as\cite{pincus}
$$\frac {d^2 \bar\phi}{dz^2}=-4\pi\ell_B [n_0 e^{-\bar\phi}-f\phi (z)], \eqno (16)$$
where $\bar\phi=\ell_B \psi$. In order to make it possible to integrate Eq. (16), we will approximate the monomer density of a brush $\phi (z)$ by the step function $\phi (z)=(N/h)\sigma\theta (h-z),$ where 
$\theta (x)=1$ for $x>0$ and 0 for $x<0$. This is a reasonable approximation because we are considering polymer brushes which are compressed because they are supporting a load, and under such circumstances, the parabolic density profile of the uncompressed brush gets flattened into a form that is not too different from the step function form given above\cite{sokoloff,milner}. Multiplying Eq. (16) by $d\bar\phi/dz$ and integrating, we get 
$$(\frac {d\bar\phi}{dz})^2=(\frac {d\bar\phi}{dz}|_{z=h})^2+$$

$$8\pi\ell_B n_0 (e^{-\bar\phi (z)}
-e^{-\bar\phi (h)} )+(8\pi\ell_B N\sigma f/h)[\bar\phi (z)-\bar\phi (h)]=$$
$$8\pi\ell_B n_0 (e^{-\bar{\phi} (z)}-1)+8\pi\ell_B N\sigma f/h [\bar{\phi} (z)-\bar{\phi} (h)], \eqno (17)$$
for z between 0 and h, using Eqs. (10-13) to simplify this expression. For $z>h$, $\bar\phi (z)=ln (cos^2 k_0(z-D/2)$, the solution described in Eqs. (10-13)\cite{safran}. Thus, Eq. (17) leads to the integral 
$$\int^{\bar\phi (z)}_{\bar\phi (h)}\frac {d\bar\phi}{[(e^{-\bar\phi}-1)+r (\bar\phi-\bar{\phi} (h)]^{1/2}}=2k_0 (z-h)$$
$$=2\alpha {z-h\over D/2-h} \eqno (18)$$
for $z<h$, where $r=K_0^2/k_0^2$, where $K_0^2=2\pi\ell_B fN/(hs^2)$, the square of the inverse screening length within a brush, where N is the number of monomers in a single polymer and s is the mean spacing of the polymers of one brush along the surface to which they are attached and $\alpha=k_0 (D/2-h)$. Evaluation of this integral allows us to obtain $\bar\phi (z)$, if we know $k_0$. The total charge between the plates consists of the sum of the charge on the polyelectrolyte brushes and the counterion charge, is zero. Applying Gauss's law, using the fact that symmetry demands that the electric fields at both plates are equal in magnitude and opposite in direction, we find that the electric field at the plates is zero, if no counterions condense on the surfaces at z=0 and z=D. Then, we must demand that at the location of the lower plate, z=0, $d\bar{\phi} (z)/dz=0.$ (It is clearly also zero at z=D by symmetry.) We can find $\bar{\phi} (z=0)$ by setting z=0 in Eq. (17), and setting $d\bar{\phi} (z)/dz|_{z=0}=0.$ (Strictly speaking, this condition is only precisely correct if the surfaces belong to thick solids, so that there can be no solution present inside these solid blocks. If this were not the case, $d\bar{\phi} (z)/dz$ would not be required to vanish at precisely z=0, but rather approximately at a short screening distance below z=0. The results are not expected to be modified qualitatively from what we will find in this section if we were to take this into account, however. In any case, this is not an important case for most applications, because there is normally no solution on the outer side of the two surfaces because there is normally solid material located there.)
\begin{figure}
\center{\includegraphics [angle=0,width=8cm]{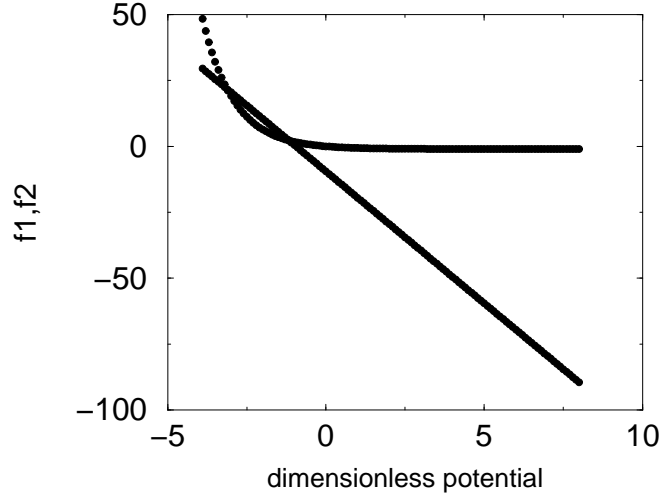}}
\caption{The functions $f1=e^{-\bar{\phi}(0)}-1$ and $f2=r(\bar{\phi} (D/2-h)-\bar{\phi}(0)$ are plotted versus the dimensionless potential $\bar{\phi} (0)$ for r=11. All quantities are dimensionless.}
\label{fig4}
\end{figure}
Using the value of $\bar{\phi} (0)$ found in this way, we may then determine $k_0$ by integrating Eq. (18) with z=0. From the definitions of $k_0$ and $K_0$ under Eqs. (10) and (18), respectively, the ratio $r=(K_0/k_0)^2$ is equal to the ratio of the unscreened charge density, $fN/(hs^2)$, of the polymer brush to $n_0$. A graphical solution for $\bar{\phi} (0)$ is illustrated in Fig. (2) for a reasonably large value of $r=(K_0/k_0)^2$. We must have $K_0/k_0\geq 1$. This is evident from the fact that (as is seen in Fig. 2) for $K_0/k_0<1$ we cannot find a solution for $\bar{\phi} (0)$ from Eq. (17) by setting $d\bar{\phi} (z)/dz|_{z=0}=0.$  We see that for any value of $r>1$, there exist two solutions for $\bar{\phi} (0)$, one with $\bar{\phi} (0)$ close to $\bar{\phi} (h)$ and one with $|\bar{\phi} (0)|>|\bar{\phi} (h)|$. Since in the large r limit, we expect the charge residing on the polymers within the polymer brushes to be highly screened, we do not expect $\bar{\phi} (0)$ and $\bar{\phi} (h)$ to differ by very much. Therefore, the solution with $\bar{\phi}(0)$ comparable to $\bar{\phi}(h)$ is the physically correct solution and the solution with $|\bar{\phi} (0)|>|\bar{\phi} (h)|$ is rejected as being an unphysical solution of the Poisson-Boltzmann equation. 
 
Eq. (18) can be integrated numerically. In order to facilitate this, following the discussion in appendix A of the second article in Ref. \cite{pincus}, we first integrate Eq. (18) by parts to eliminate the singularity in the integrand, giving 
$$\frac{tank_0 (D/2-h)}{K_0^2/k_0^2-1-tan^2 k_0 (D/2-h)}$$
$$-\int_{\bar{\phi}(h)}^{\bar{\phi}(0)}d\bar{\phi}
{e^{-\bar{\phi}}[e^{-\bar{\phi}}-1+r(\bar{\phi}-\bar{\phi}(h)]^{1/2}\over (r-e^{-\bar{\phi}})^2}=k_0 h. \eqno (19)$$
Eq. (19) may also be written as 
$${h\over D/2-h}={\alpha tan\alpha\over \beta^2-\alpha^2 (tan^2\alpha+1)}$$
$$-\alpha^{-1}\int_{\bar{\phi}(h)}^{\bar{\phi}(0)}d\bar{\phi}
{e^{-\bar{\phi}}[e^{-\bar{\phi}}-1+r(\bar{\phi}-\bar{\phi}(h)]^{1/2}\over (r-e^{-\bar{\phi}})^2}, \eqno(20)$$
where $\alpha=k_0 (D/2-h)$ and $\beta=K_0 (D/2-h)$. 
Several solutions of Eq. (20) are given in table I. In each case, we chose the largest value of $\alpha$ for which there is a solution to Eq. (18) with z=0 for which $d\bar{\phi} (z)/dz|_{z=0}=0$. It was found to be difficult to find a solution for $D/2-h<h$. It is clear that it must be possible to have solutions to Eq. (20) with $D/2-h<h$. It is clear, however, that in order to have a solution to Eq. (18) with  $D/2-h<h$, the left hand side of this equation must be greater than 1, unless $\alpha$ is small, and we already saw for small values of $\alpha$, Eq. (20) gives values of (D/2-h)/h which are greater than 1. For reasonably large values of r, the only way that this can occur is if there is a large contribution to the integral from $\bar{\phi}$ close to $\bar{\phi} (0)$, where the integrand has an integrable singularity. Let us then examine the contribution to the integral in Eq. (18) for this part of the range of integration. To accomplish this, we write $\bar{\phi}=\bar{\phi} (0)+\phi'$ and perform the integral over a range of $\phi'$ which is small compared to $\bar{\phi} (h)-\bar{\phi} (0)$. Then this contribution to the integral can be performed by expanding the function under the radical sign in the denominator of the integrand in a Taylor series to second order in $\phi'$, giving $f(\bar{\phi})=e^{-\bar{\phi}}-1+r(\bar{\phi}-\bar{\phi} (h))\approx df/d\bar{\phi}|_{\bar{\phi}=\bar{\phi} (0)} \phi'+(1/2)d^2f/d\bar{\phi}^2|_{\bar{\phi}=\bar{\phi} (0)}\phi'^2$. Using this, the contribution to the integral in Eq. (18) for $\bar{\phi}$ close to $\bar{\phi} (0)$ can be written as  
$$\Gamma^{-1/2}\int_0^{\phi_1} {d\phi'\over [\phi'^2+\delta\phi']^{1/2}}\approx \Gamma^{-1/2}[ln(2(\phi_1))-ln(\delta/2)], \eqno (21)$$
where $\Gamma=(1/2)d^2f/d\bar{\phi}^2|_{\bar{\phi}=\bar{\phi} (0)}$, $\Gamma\delta=df/d\bar{\phi}|_{\bar{\phi}=\bar{\phi} (0)}$ and $\delta<<\phi_1<<\bar{\phi} (h)-\bar{\phi} (0)$. From the definition of $\delta$, we see that $\delta=e^{\bar{\phi} (0)}df/d\bar{\phi}|_{\bar{\phi}=\bar{\phi} (0)}$ and since $f(\bar{\phi})=f1-f2$, defined in Fig. 2, we see that $\delta$ becomes smaller and smaller as the curves f1 and f2 in Fig. 2 become more and more nearly tangent at $\bar{\phi}=\bar{\phi} (0)$. From Eq. (21), we see that we can make $(D/2-h)/h$ as small as we wish by making $\delta$ smaller [and hence making the curves closer to being tangent at their point of intersection at $\bar{\phi} (0)$]. For example, for the value of $\beta=1.17$, which we argued earlier to be appropriate for the polyelectrolyte brushes studied in Ref. \cite{raviv}, we find that the curves will be nearly tangent for $\alpha$ slightly larger than 0.65. This value of $\alpha$ corresponds to $n_0=[0.65/(2\pi)]/[\ell_B (D/2-h)^2]=0.7\times 10^{26}m^{-3}$.  
Thus, we conclude that the density of counterions residing between the tops of the two brushes is sufficient to provide enough osmotic pressure to support a load of $k_B T n_0=0.28\times 10^6Pa$ for the polyelectrolyte brushes studied in Ref. \cite{raviv}.

The degree of compression can be estimated from the discussion in Ref. \cite{pincus}. If the brushes are being pushed together with a compressional force or load per unit area P, we have by the scaling arguments of Ref. \cite{pincus} that 
$$P\approx {fNk_B T\over s^2 h}-{hk_B T\over Na^2 s^2}, \eqno (22)$$
Setting P equal to the osmotic pressure in the interface region between the two brushes, $k_B T n_0$, we may write the solution of Eq. (25) for h as 
$$h=f^{2/3} Na[(1+f^{-1}(as^2 n_0/2)^2)^{1/2}-as^2n_0/(2f^{1/2})].  \eqno (23)$$
If we use the following values for the parameters: $n_0=10^{25}m^{-3}$, $a=2.5A^o$ and $s=40A^o$, we find that h is compressed to about 42 percent of its maximum value of $f^{1/2}Na$ 

We will see in the next section that there also exists a solution with r comparable to 1, if the charge density of the unscreened polymer brushes $fN/(s^2 h)$ is not much larger than $(\pi/2)/[(D/2-h)^2\ell_B]$. 

\begin{table}

\noindent\caption{Results of Calculations} based on Eq. (20)
\begin{tabular}{||c|c|c|c|c||}
\hline\hline
$K_0 (D/2-h)$&$k_0 (D/2-h)$&(D/2-h)/h\\

\hline\hline

1.17& 0.6&2.30\\ \hline

0.70& 0.45&1.40\\ \hline

0.30&0.35&1.05\\ \hline

1.17&0.20&37.7\\ \hline

\end{tabular}

\end{table}
 
\section{Load Carrying Ability of Polyelectrolyte Brushes with Purely Electrostatic Interactions}

Zhulina and Borisov have published an analytic solution of the mean field theory of polyelectrolyte brushes\cite{zhulina}, assuming that the only interaction between the polymers in a brush is that which results from electrostatic interaction and entropic interactions resulting from the counterions. The main conclusion of this treatment relevant to the present work is that this model predicts that under such circumstances, for highly charged polyelectrolyte brushes practically all of the counterions lie inside the brushes. The reason for this is that if we assume that there are no hard core interactions between the monomers in the brushes (i.e., there are only electrostatic and counterion entropic interactions), the mean field potential is identically equal to the electrostatic potential energy of the monomers, which is equal to $fe\psi ({\bf r})$. Since the mean field potential must have the form $A-Bz^2$ \cite{milner}, the net charge density in units of the electron charge e (due to both monomer and counterion charge) is given by $-(\epsilon/fe^2) d^2\psi/dz^2=(2\epsilon/fe^2)B=\pi^2/(8f\ell_B a^2 N^2)$, a constant. Since the product of this charge density and the volume of the brush must be equal in magnitude to the total amount of counter ion charge outside the brush (in units of e), for monovalent counterions, the number of counterions outside two brushes in contact with each other is equal to $\pi^2 hs^2 (8f\ell_B a^2 N^2)$. Then, clearly for highly compressed (i.e., $h<<f^{1/2}Na$ and obviously $s<<f^{1/2}Na$) and highly charged polymers (i.e., $f\approx 1$), this model predicts that there are practically no counterions outside the brushes. This model predicts that for highly compressed polyelectrolyte brushes, i.e., brushes with heights $h<<f^{1/2} Na$ (see Fig. 8 in Ref. \cite{zhulina}), the density profiles will approximate a step function and the tops of the two brushes will never be in contact (i.e. $D>2h$). Hence, there will always be a thin interface region free of monomers separating the two brushes, but the thickness of this region will be negligibly small (i.e., less than $1A^o$), in order to support a load as high as $10^5 Pa$. This would imply on the basis of the arguments given in section 2, that highly compressed polymer brushes should exhibit a good deal of friction due to entanglement of blobs belonging to one brush in the second brush. (How the scaling arguments of Ref. \cite{sokoloff} are modified when the polymer interactions are replaced due to electrostatic, rather than hard core, interactions will be discussed in a future publication\cite{sokoloff1}. 

Hence, the calculations presented in the previous paragraph must give results consistent with Ref. \cite{zhulina} when $(D-2h)/D<<1$. In order to make contact between Ref. \cite{zhulina} and the Poisson-Bltzmann equation calculations  presented in the previous section, let us do such a calculation for an example with $(D-2h)/D<<1$ with some reasonable parameters. Consider $(K_0 D/2)^2=2\pi\ell_B(2fN/(\pi^{1/2} hs^2)D^2$, which for $\ell_B=7A^o$, $fN=10^3 A^o$, $s=10^2A^o$ and $h\approx D/2\approx 10^2 A^o$, we get $K_0 D=15.8$. The solution to Eq. (18) for $(D-2h)/D=0.001$ gives $k_0 D/2=15.79$, which implies that $n_0$ is only slightly smaller than the unscreened monomer charge density inside a brush and $\bar{\phi} (h)-\bar{\phi} (0)=0.262$, which implies that $\bar{\phi} (z)$ varies by a relatively small fraction as z varies from 0 to h, which is consistent with the calculation of Zhulina and Borisov\cite{zhulina}. An alternative way to view this is the following: In Zhulina and Borisov's solution of mean field theory\cite{zhulina}, highly charged polyelectrolyte brushes have practically no net charge inside the brushes, meaning that the counterion density inside the brush is nearly equal to the unscreened polymer charge density, as shown above. Consequently since the counterion density is continuous as we cross over from inside to outside the brush, $k_0$ is only slightly less than $K_0$. Since $(K_0/k_0)^2$ is equal to the ratio of $n_0$, the counterion density midway between the brushes, to the charge density of the unscreened polymers inside the brushes, we may write $(K_0/k_0)^2=1+\Delta n/n$, where $\Delta n$ is the net charge density inside the brushes, and n is the unscreened polymer charge density. Since $\Delta n/n<<1$, Eq. (18) reduces to 
$$tank_0 (D/2-h)\approx k_0 h(\Delta n/n) \eqno (24)$$
to lowest order in $\Delta n/n$. Since the right hand side of Eq. (18) is much less than 1, the tangent may be approximated by its argument, and hence we find from Eq. (24) that 
$$D/2-h\approx h(\Delta n/n). \eqno (25)$$
From the discussion above, we know that $\Delta n=\pi^2/(8f\ell_B a^2 N^2)$. The unscreened polymer charge density is $fN/(s^2 h)$. Using the same values for f, $\ell_B$, a, N and h as were used above we estimate that 
$$D/2-h)/(D/2)\approx 0.4\times 10^{-4}. \eqno (26)$$

Thus, there exist two treatments of highly charged compressed polyelectrolyte brushes, which appear to give opposite results. One predicts that polyelectrolyte brushes should be able to slide with their load supported by osmotic pressure in a region separating the tops of the brushes which is relatively free of monomers of sufficient thickness to prevent polymer entanglement of the type that was proposed in ref. \cite{sokoloff} to give relatively high friction compared to the viscous friction considered in section II and one which predicts that the load will be supported by osmotic pressure in a region separating the brushes which is extremely thin. The latter would imply that there should always be a good deal of friction due to blob entanglement when $h<<f^{1/2} Na$, on the basis of Ref. \cite{sokoloff}. Let us now explain this apparent contradiction. When the polymer charge is relatively high (i.e., f is close to 1), Zhulina and Borisov solution of mean field theory can only be valid if the brushes are somewhat compressed, because if the brushes are as stretched as far as Pincus's treatment\cite{pincus} implies, the elastic free energy of the polymers used in Ref. \cite{zhulina} is not valid for highly extended polymers. When the brushes are compressed to heights comparable to the heights that they would have if the polymers were neutral, the hard core repulsion of the monomers will play a role comparable to that of the electrostatic interactions. Although mean field theory requires that the sum of the hard core interaction and electrostatic interaction potentials be equal to $A-Bz^2$, with the constant B having the value given earlier, the electrostatic interaction potential alone need not have this form. In fact, when the brushes are highly compressed (compared to the nearly fully extended brush height bound in Ref. \cite{pincus}) down to a height comparable to what one finds for neutral brushes, the hard core potential might dominate. In such a case, the electrostatic potential is certainly not constrained to have the above form. Rather, it will be determined completely by the solution of the Poisson-Boltzmann equation using a monomer density profile, determined primarily by the hard core interaction contribution to mean field theory, in which case a solution like the one found in section III is likely to be valid. Whether or not the polyelectrolyte brushes are so highly compressed that hard core interaction plays a dominant role, the density profile of polymer brushes formed from finite length polymers will have significant deviations from mean field theory in the form of tails of length $\ell<<h$ at the edges of the brushes. This will certainly require that the mean field potential differ from the parabolic from required by mean field theory\cite{milner} at the edges of the brushes (i.e., near z=h) over a region of thickness of the order of $\ell$. Then the electric charge density obtained by differentiating the electrostatic potential twice will have a charge density which is much larger than that found be Zhulina and Borisov\cite{zhulina} in this region, resulting in a considerably larger net charge of the polymer brush than that found in Ref. \cite{zhulina}. 

\section{Effects of Excess Salt}

When there is excess salt present in the solvent, Eq. (16) gets replaced by\cite{safran} 
$$\frac {d^2 \bar\phi}{dz^2}=-4\pi\ell_B [n_s (e^{-\bar\phi}-e^{+\bar\phi})-f\phi (z)], \eqno (27)$$
where $n_s$ represents the salt concentration and the second exponential term on the right hand side represents the contribution of ions with the same charge as the brushes to the ionic charge between the plates. For the case of no excess salt, described by Eq. (16), $\bar{\phi}(z)$ was taken to be zero at z=D/2. In contrast, for the case of excess salt, described by Eq. (20), $\bar{\phi}(z)$ is usually taken to be zero well outside the two surfaces. For a low concentration of excess salt, the conditions under which Eq. (20) takes the same form as Eq. (16) will be examined using simple physical arguments, which give results which are identical to those obtained in appendix A from the exact solution to the Poisson-Boltzmann equation of the second article in Ref. \cite{pincus}. In order to accomplish this, let us write $\bar{\phi}(z)$ as $\bar{\phi}(z)=\phi_0+\phi'(z)$ where $\phi_0=\bar{\phi}(D/2)$ and $\phi' (z)$ is zero at z=D/2. Then, we can make Eq. (27) look like Eq. (16), if we identify $n_0$ with $n_s e^{-\phi_0}$. Then we can write Eq. (27) as 
$$\frac {d^2 \phi'}{dz^2}=-4\pi\ell_B [n_0 (e^{-\phi'}-(n_s/n_0)e^{+\phi'})-f\phi (z)]. \eqno (28)$$
When $n_s/n_0=e^{\phi_0}<<1$, Eq. (16) is definitely a good approximation to the problem, and we are justified in treating the system as one without excess salt. To determine the conditions under which Eq. (16) is a good approximation, we solve Eq. (16), in order to determine $n_0$ as described earlier in this section, and determine $\phi_0$ from $n_0=n_s e^{-\phi_0}$, and use $\phi_0$ and the solution of Eq. (16) for $\bar{\phi} (z)$, which we identify with $\phi' (z)$ (which is the approximate solution to Eq. (28)) to determine the conditions under which we may neglect $e^{\bar{\phi} (z)}$ compared to $e^{-\bar{\phi} (z)}$. (Remember that $\bar{\phi} (z)$ is negative.) Since it is easily seen from Eq. (10) that $e^{-\phi' (z)}$ is  significantly greater than 1 over much of the range of z from 0 to h for high density polymer brushes, all that is required in order to neglect $e^{\bar{\phi} (z)}$ is that $n_s/n_0$ be of order unity, which is already satisfied for the 0.1 M salt concentration (or $0.6\times 10^{26}m^{-1}$) typical of living matter. It is demonstrated in appendix A that these results follow directly from the exact solution of the Poisson-Boltzmann equation with excess salt present\cite{pincus}.

\section{Conclusions}

It has been shown using a modified version of the mean field theory of Miklavic Marcelja\cite{miklavic} for polyelectrolyte polymer brushes, which uses the non-linear Poisson-Boltzmann equation, that it should be possible for osmotic pressure due to the counterions to support a reasonably large load (about $10^6 Pa$) with the tops of the brushes sufficiently far apart to prevent entanglement of polymers belonging to the two brushes, which has been argued to account for most of the friction. This load carrying ability is argued to persist in the presence of an amount of added salt comparable to that found in living matter. Significant additional salt, however, provides screening which reduces the load carrying ability of polyelectrolyte brushes. Using counterions of higher valence will also not improve the load carrying because it will actually reduce the net charge 
on the polymers, by causing more counterions to condense\cite{manning}. This will in turn reduce the counterion concentration in solution. The load carrying ability of the brushes could be improved by using a solvent with a higher dielectric constant, which would reduce the value of the Bjerrum length, which appears in the denominator of the expression for the counterion osmotic pressure [Eq. (15)]. Using denser brushes or better solvents, which increase the ratio of brush height to polymer radius of gyration would improve the load carrying ability, 
since the minimum thickness of the interface region between the polymer brushes 
which avoids entanglements that lead to static friction, is given approximately by the quantity $\xi$ under Eq. (8) is proportional to $h^{-1/3}$. Making this region less thick increases the counterion osmotic pressure, as Eq. (15) shows that it is inversely proportional to the square of the interface region thickness. 

\section {Appendix A: The high and Low Salt Concentration Limits of the Exact Solution of the Poisson-Boltzmann Equation}

The exact solution to the Poisson-Boltzmann equation [i.e., Eq. (20)] given in the second article in Ref. \cite{pincus} for the electrostatic potential in the interface region between two polymer brushes in terms of elliptic functions, which are close to each other but not touching, is 
$$\bar{\phi} (z)=2arcsinh[\frac{sinh(\bar{\phi}(D/2)/2)}{cn(K\bar{z} cosh(\bar{\phi} (D/2)/2),k)}], \eqno (1A)$$
where $\bar{z}=D/2-h$ and $K=(8\pi\ell_B n_i)^{1/2}$, where $n_i$ is the salt concentration, $c_n (x,k)$ is an elliptic function and k, the standard elliptic function parameter k \cite{jahnke}, is given by $k=[\cosh(\bar{\phi} (D/2)/2)]^{-1}$.  First, let us consider the limit as $\bar{\phi} (D/2)$ approaches zero. From Eq. (1A) we find that 
$$\bar{\phi} (z)\approx \frac{\bar{\phi} (D/2)}{cn(K\bar{z},1)}=\bar{\phi} (D/2) cosh[K(D/2-z)] \eqno (2A)$$
since $cn(x,k=1)=sech(x)$. This is the solution in the Debye-Huckel approximation. 

Now let us consider the limit as $|\bar{\phi} (D/2)|$ approaches infinity. Since $cosh(\bar{\phi} (D/2)$ approaches infinity, k in Eq. (1A) approaches zero. It is easily shown that cn(u,k=0)=cos(u). Then, from Eq. (1A) we obtain 
$$sinh(\bar{\phi}/2)\approx {sinh(\bar{\phi}(D/2)/2)\over cn(Kzcosh(\bar{\phi} (D/2)/2),0)}$$
$$={sinh(\bar{\phi}(D/2)/2)\over cos(Kzcosh(\bar{\phi} (D/2)/2))},$$
which becomes for $|\bar{\phi}(D/2)|>>1$, 
$$\bar{\phi} (z)\approx \bar{\phi} (D/2)+ln|cos^2 [(K(z-D/2)cosh(\bar{\phi}(D/2)/2)]|. \eqno (3A)$$
Since as $|\bar{\phi} (D/2)|$ approaches infinity, $cosh(\bar{\phi}(D/2)/2)$ is approximately equal to $(1/2)e^{|\bar{\phi} (D/2)|/2}$,  $Kcosh(\bar{\phi} (D/2)/2)$ becomes $[2\pi\ell_B n_0]^{1/2}$, with $n_0=n_s e^{|\bar{\phi}(D/2)|}=n_s e^{-\bar{\phi} (D/2)}$. The latter result follows from the fact that since the counterions must have lower potential energy between the plates than outside the plates, $\bar{\phi} (D/2)<0$.

\section{Acknowledgment}
I wish to thank R. Tadmor for many discussions of his work and P. L. Hansen of MEMPHYS for insightful discussions.



\end{document}